\documentclass[preprint2]{aastex}

\shorttitle{New models of partially ionized disks}
\shortauthors{C. Terquem}

\begin{document}

\title{New composite models of partially ionized protoplanetary disks}

\author{Caroline E. J. M. L. J. Terquem\altaffilmark{1} }
\affil{Institut d'Astrophysique de Paris, UMR 7095 CNRS, 
Universit\'e Pierre \& Marie Curie Paris 06, 98 bis bd Arago, 75014
Paris, France}

\email{caroline.terquem@iap.fr}

\altaffiltext{1}{Institut Universitaire de France}

\begin{abstract}

We study an accretion disk in which three different regions may coexist:
MHD turbulent regions, dead zones and gravitationally unstable regions.
Although the dead zones are stable, there is some transport due to the
Reynolds stress associated with waves emitted from the turbulent layers.
We model the transport in each of the different regions by its own
$\alpha$ parameter, this being 10 to $10^{3}$ times smaller in dead
zones than in active layers.  In gravitationally unstable regions,
$\alpha$ is determined by the fact that the disk self--adjusts to a
state of marginal stability.  We construct steady--state models of such
disks.  We find that for uniform mass flow, the disk has to be more
massive, hotter and thicker at the radii where there is a dead zone.  In
disks in which the dead zone is very massive, gravitational
instabilities are present.  Whether such models are realistic or not
depends on whether hydrodynamical fluctuations driven by the turbulent
layers can penetrate all the way inside the dead zone.  This may be more
easily achieved when the ratio of the mass of the active layer to that
of the dead zone is relatively large, which in our models corresponds to
$\alpha$ in the dead zone being about 10\% of $\alpha$ in the active
layers.  If the disk is at some stage of its evolution not in
steady--state, then the surface density will evolve toward the
steady--state solution.  However, if $\alpha$ in the dead zone is much
smaller than in the active zone, the timescale for the parts of the disk
beyond a few AU to reach steady--state may become longer than the disk
lifetime.  Steady--state disks with dead zones are a more favorable
environment for planet formation than standard disks, since the dead
zone is typically 10 times more massive than a corresponding turbulent
zone at the same location.

\end{abstract}



\section{Introduction}

Protoplanetary disks are believed to have a complex structure.  In the
parts which are ionized enough ({\em active} zones), the magnetic field
couples to the gas and the magnetorotational instability (MRI; Balbus~\&
Hawley~1991) develops, leading to turbulence and angular momentum
transport.  Some parts however are too cold and dense for the ionization
to reach the level required for MHD turbulence to be sustained.
However, in these so--called {\em dead} zones, there may still be some
low level of transport due to Reynolds or non turbulent Maxwell stresses
(Fleming \& Stone 2003, Turner \& Sano 2008).  Finally, parts of the
disk may be dense enough that gravitational instabilities develop,
redistributing mass and angular momentum in such a way that the disk
settles into a state of marginal stability.  To date, there is no
numerical simulations of such global disks.  Global three dimensional
MHD simulations so far have modeled stratified and fully turbulent
disks, but with no dead zones (Fromang~\& Nelson~2006).

It is of interest to study whether non--uniform disks composed of
different regions as described above can be in steady--state or not.
Previous studies (Gammie 1999, Armitage et al. 2001) assumed that no
transport at all took place in the dead zone.  Therefore mass could only
pile--up there until gravitational instabilities would develop.  If the
level of transport is non zero in the dead zone however, the disk may be
able to adjust to reach a steady--state.  It is this possibility that we
investigate in this paper.

In determining the characteristics of the disk vertical strcuture,
we ignore reprocessing of the stellar radiation.  Only heating
associated with the different (magnetic and hydrodynamic) stresses is
considered.  Reprocessing has been shown to be important when the disk
surface is flared (Kenyon ~\& Hartmann 1987, Chiang \& Goldreich 1997,
D'Alessio et al. 1998, Dullemond et al. 2001).  It provides an extra
heating source that dominates the disk beyond a few AU and has important
consequences on the structure of the dead zone (Matsumura \& Pudritz
2006).  However, in the models we present below, the parts of the disk
beyond 1~AU are shielded from the stellar radiation by the inner parts.
These models, which do not include reprocessing, are therefore
self--consistent.

The plan of the paper is as follows.  In section~\ref{sec:modeling} we
describe the way we model the transport in a disk with turbulent
regions, dead zones and gravitationally unstable regions.  In
section~\ref{sec:vertical} we present the equations governing the disk
vertical structure and describe how they are solved.  In
section~\ref{sec:results} we present the results of the calculations,
and discuss them in section~\ref{sec:discussion}.  We find that
steady--state solutions exist and correspond to disks which are thicker,
hotter and more massive than disks without dead zones.

\section{Modeling of the disk transport}
\label{sec:modeling}

Protoplanetary disks are almost certainly magnetized, as they form from
molecular clouds which are observed to contain a magnetic field.  If the
magnetic pressure is smaller than the thermal pressure, then the parts
of the disk where the field couples to the gas are prone to the MRI.
Its nonlinear developement is turbulence that transports angular
momentum outward and therefore enables accretion (Hawley et al. 1995,
Brandenburg et al. 1995).  Recent numerical simulations of the MRI in
shearing--boxes have shown that the level of transport depends on the
Prandtl number (ratio of the viscosity to the resistivity; Fromang et
al. 2007, Lesur~\& Longaretti~2007).  The simulations, which have not
converged yet, also show that the turbulence is not sustained when the
net magnetic flux is zero and for the range of parameters investigated
(relatively large Prandtl numbers).
%For the range or parameters investigated (relatively low Reynolds
%numbers), and if there is no net magnetic flux through the box, the
%simulations, which have not converged yet, show that the turbulence is
%not sustained for Prandtl numbers under some critical value that
%decreases with the Reynolds number (Fromang et al. 2007).  When there is
%a net flux, the turbulence is always sustained but the transport
%coefficient decreases when the Prandtl number decreases (Lesur~\&
%Longaretti~2007).  
It is reasonable to suppose that the magnetic field in protoplanetary
disks does not vary very much on the scale of a few scale heights.  For
this reason, and if these disks were to be described locally by a
shearing box, it would be more appropriate to assume a finite net flux.
In the parts of the disk that are coupled to the field, we will
therefore assume that turbulence is sustained and quantify the angular
momentum transport using the parametrization of Shakura~\&
Sunyaev~(1973), with $\alpha_T$ being the ratio of the stress to the
pressure (the subscript 'T' refers to turbulence).  It has been shown
that this prescription, which is formally valid only if the disk evolves
under the action of a viscosity, can be used to describe the global
properties and evolution of MHD turbulent disks, although details and
stability issues cannot be studied within this framework (Balbus~\&
Papaloizou~1999).  Given that $\alpha_T$ as measured in the simulations
depends on the Prandtl number, and that Prandtl numbers typical of
protoplanetary disks are way too small to be investigated numerically,
we will assign values to $\alpha_T$ based on observations.  The
diffusion timescale is indeed limited by the disk lifetime, which is
observed to be of a few $10^6$ years.  This implies $\alpha_T \sim
10^{-3}$--$10^{-2}$ (at least in the outer parts of the disk which are
fully turbulent).

In protoplanetary disks, the minimum ionization fraction which is
required for MHD turbulence to be initiated and sustained is on the
order of $10^{-13}$ on scales of 1~AU (e.g., Balbus~\& Hawley~2000).
Although very small, this level of ionization is not reached in some
parts of the disks, which are relatively cold and dense.  The extent of
the dead zone has been computed by several authors, taking into account
ionization by cosmic rays (Gammie 1996, Sano et al.~2000 who also
included radioactivity) and X--rays (Igea~\& Glassgold 1999, Fromang et
al. 2002, Matsumura \& Pudritz 2003 who also included cosmic rays
and radioactivity).  In this paper, we will denote by $\Sigma_T$ the
surface density of the active layer, i.e. the surface density beyond
which ionizing photons cannot penetrate.  Depending on the energy of
these photons, $\Sigma_T$ may vary in the range 10--100~g~cm$^{-2}$
(Gammie 1996, Fromang et al. 2002).  Note that the ionization in
the disk inner parts is thermal, with the most abundant ions being
Na$^+$ and K$^+$.  For the ionization fraction to be larger than
$10^{-13}$, the temperature has to be higher than some value $T_i$ which
is close to $10^{3}$~K over a large range of densities (Balbus \& Hawley
2000).
%Recently, Inutsuka~\& Sano (2005) have questioned the existence of dead
%zones.  They argue that, once the turbulence is initiated, the electrons
%that maintain the current and hence the magnetic field have velocities
%large enough to keep through collisions the degree of ionization at the
%level required for the turbulence to be sustained.  Whether this process
%works or not depends very sensitively on the energetics of the
%electrons, and a detailed analysis of the energy losses has to be
%carried out before the existence of dead zones can be ruled out.
%Therefore, in this paper, we will assume that the parts of the disks in
%which X--rays and/or cosmic rays cannot penetrate enough to ionize the
%gas at the required level and which are too cold for thermal ionization
%to be significant are not turbulent.
There is evidence from numerical simulations that some level of
transport can be present in a dead zone that is sandwiched in between
turbulent layers (Fleming~\& Stone~2003), as the turbulence drives
hydrodynamic waves that propagate in the dead zone and are associated
with a Reynolds stress.  More recently, Turner~\& Sano (2008) have also
proposed that magnetic field be able to diffuse from the active layers
into the dead zone.  Although the resistivity there is too high for MHD
turbulence to be sustained, there is a large scale magnetic stress
associated with the field that drives accretion.  (If dust grains are
present in the disk however, the magnetic field may not able to diffuse
down to the disk midplane at all radii.)  The transport and energy
deposition in the dead zone due to these stresses are non local.
However, for illustrative purposes, we will parameterize this transport
with a Shakura~\& Sunyaev parameter that we will denote $\alpha_D$
(where the subscript 'D' refers to dead zone).  Fleming~\& Stone~(2003)
found that the Reynolds stress in the dead zone is about 10\% of the
Mawell stress in the active layers.  Normalized by the thermal pressure,
this translates into $\alpha_D$ being between 0.07 and 0.3 times
$\alpha_T$.  However, in their simulations, the mass of the dead zone
was only a few times that of the active layer.  We might expect lower
values of the Reynolds stress deep into the dead zone when the mass of
the active layer is significantly smaller than that of the active zone,
which is the case in the models we present below.  Therefore, we will
take $\alpha_D$ in the range $10^{-3}\alpha_T$ -- $10^{-1} \alpha_T$.

In the parts of the disk which are dense and cold, gravitational
instabilities develop.  Spiral density waves grow and transport angular
momentum outward and mass inward.  The level of saturation is determined
by the fact that the disk settles into a state of marginal stability,
with the Toomre parameter $Q$ being on the order of 1.5 (Laughlin~\&
Bodenheimer~1994, Lodato~\& Rice~2004).  It was shown by Balbus~\&
Papaloizou~(1999) that in general, in a gravitationally unstable disk,
the transport is non local and cannot be described using the viscous
disk theory.  However, these authors pointed out that a local transport
model might apply if the disk maintains itself in a state of marginal
stability, which is the case considered here.  In the regions where the
disk is gravitationally unstable, we will therefore use the Shakura~\&
Sunyaev prescription and will adjust the transport parameter, that we
will denote $\alpha_G$ (where the subscript 'G' refers to gravity), such
as to have $Q \sim 1.5$.

It is important to keep in mind that here the disk is modeled in a very
schematic way.  We believe this does not affect the gross features of
the models we present below, although the details should be considered
with caution.

\section{Disk vertical structure}
\label{sec:vertical}

We consider a system of cylindrical coordinates $(r,\varphi, z)$ based
on the central star, with $z=0$ being the disk midplane. $P$ is the
pressure, $\rho$ is the mass density per unit volume, $T$ is the
temperature, $\Omega$ is the angular velocity, $\nu$ is the 'enhanced'
kinematic viscosity associated with $\alpha_T$ in turbulent layers,
$\alpha_D$ in dead zones and $\alpha_G$ in gravitationally unstable
regions, and $\kappa$ is the opacity, which in general depends on both
$\rho$ and $T$. The thin--disk approximation is used throughout (and
checked to be valid {\em a posteriori)}, so that $\Omega^2 =
GM_{\star}/r^3$, $M_{\star}$ being the mass of the central star and $G$
the gravitational constant.

The disk vertical structure is described by the equation of vertical
hydrostatic equilibrium:

\begin{equation}
\frac{1}{\rho} \frac{\partial P}{\partial z} = - \Omega^2 z ,
\label{dPdz}
\end{equation}

\noindent  together with the energy equation, which states that the
rate of energy removal by radiation is locally balanced by the rate of
energy production by viscous dissipation:

\begin{equation}
\frac{\partial F}{\partial z} = \frac{9}{4} \rho \nu \Omega^2 
= \frac{9}{4} \alpha \Omega P ,
\label{dFdz}
\end{equation}

\noindent where $F$ is the radiative flux of energy through a surface of
constant $z$.  To write the last term of this equation, we have used
$\nu=\alpha c_s^2/\Omega$, where $c_s=(P/\rho)^{1/2}$ is the isothermal
sound speed and $\alpha=\alpha_T$ in active layers, $\alpha=\alpha_D$ in
dead zones and $\alpha=\alpha_G$ in the gravitationally unstable parts
of the disk.  Therefore $\alpha$ is a function of both $z$ and $r$.  The
flux $F$ is given by:

\begin{equation}
F = \frac{- 16 \sigma T^3}{3 \kappa \rho}
\frac{\partial T}{\partial z} ,
\label{dTdz}
\end{equation}

\noindent with $\sigma$ being the Stefan--Boltzmann constant.  In
principle, equation~(\ref{dTdz}) is valid only at radii where the disk is
optically thick.  However, when the disk is optically thin, i.e. when
$\kappa \rho$ integrated over the disk thickness is small compared to
unity, the temperature gradient given by equation~(\ref{dTdz}) is small,
so that the results we get are consistent in that case also.  
To close the system of equations, we adopt the equation of state of an
ideal gas:

\begin{equation}
P = \frac{\rho k T}{2 m_H} ,
\label{state}
\end{equation}

\noindent where $k$ is the Boltzmann constant and $2 m_H$ is the mass of
the hydrogen molecule, which is the main component of protostellar disks
at the temperatures and densities of interest here.

\subsection{Boundary conditions}

We have to solve three first order ordinary differential equations for
the three variables $F$, $P$ (or equivalently $\rho$), and $T$ as a
function of $z$ at a given radius $r$.  Accordingly, we need three
boundary conditions at each $r$.  These have been described in detail in
Papaloizou~\& Terquem~(1999), so here we just recall briefly their
expression.  We denote with a subscript $s$ values at the disk surface.
The flux at the surface is given by:

\begin{equation}
F_s = \frac{3}{8 \pi} \dot{M} \Omega^2 ,
\label{Fs}
\end{equation}

\noindent where $\dot{M} = 3 \pi \langle{\nu}\rangle \Sigma$, with
$\Sigma = \int_{-H}^H \rho dz$ being the disk surface mass density and
$\langle \nu \rangle = \int_{-H}^H \rho \nu dz/ \Sigma$ being the
vertically averaged viscosity.  If the disk is in a steady state,
$\dot{M}$ does not vary with $r$ and is the constant accretion rate
through the disk. $H$ and $-H$ are the upper and lower boundaries of the
disk, respectively.  The surface pressure is given by:

\begin{equation}
P_s = \frac{\Omega^2 H \tau_{ab}}{\kappa_s} ,
\label{Ps}
\end{equation}

\noindent where $\tau_{ab}$ is the optical depth above the disk.  Since
we have defined the disk surface such that the atmosphere above the disk
is isothermal, we have to take $\tau_{ab} \ll 1$.  Providing this is
satisfied, the results do not depend on the value of $\tau_{ab}$ we
choose (see Papaloizou~\& Terquem~1999).  Finally, the surface
temperature satisfies the following equation:

\begin{equation}
2 \sigma \left( T_s^4 - T_b^4 \right) - \frac{9 \alpha_s k T_s \Omega}{8
\mu m_H \kappa_s} - \frac{3}{8 \pi} \dot{M} \Omega^2 = 0 .
\label{Ts}
\end{equation}

\noindent Here the disk is assumed immersed in a medium with background
temperature $T_b$, so that the surface temperature remains finite.

\subsection{Model Calculations}
\label{sec:calcul}

We look for steady--state solutions, so that $\dot{M}$ does not depend
on $r$.  At a given $r$ we then solve equations~(\ref{dPdz}),
(\ref{dFdz}) and~(\ref{dTdz}) with the boundary conditions~(\ref{Fs}),
(\ref{Ps}) and~(\ref{Ts}) to find the dependence of the state variables
on $z$.  The (Rosseland) opacity is taken from Bell~\& Lin (1994). This
has contributions from dust grains, molecules, atoms and ions.  It is
written in the form $\kappa=\kappa_i \rho^a T^b$ where $\kappa_i$, $a$
and $b$ vary with temperature.  To carry out the integration, we start
from an estimated value of $H$ and iterate until the condition $F=0$ at
$z=0$ is satisfied (see details in Papaloizou~\& Terquem~1999).  The
integration starts at $z=H$, where we have $\alpha=\alpha_T$, assuming
the disk is not gravitationally unstable, and proceeds all the way down
to $z=0$.  At each value of $z$, we compute $\Sigma(z)=\int_z^{H}
\rho(z') dz'$ and compare it with $\Sigma_T$.  When
$\Sigma(z)>\Sigma_T$, and if $T(z)<T_i$, we fix $\alpha=\alpha_D$.  Once
the vertical integration is finished, we calculate the Toomre parameter
$Q=c_s \Omega / \pi G \Sigma$, where $\Sigma$ is now the total surface
density and $c_s$ is evaluated at the midplane.
% As $\Sigma$ is larger in the parts of the disk where there is a dead 
% zone, $Q$ is smaller there.  
When $Q<1.5$, we increase $\alpha$, that we now call $\alpha_G$,
slightly, assume that it is now independent of $z$, and redo the
integration.  We iterate the calculation, increasing $\alpha_G$ at each
step, until $Q$ becomes larger than 1.5.  As the step with which we
increase $\alpha_G$ has a finite size, we usually end up with a value of
$Q$ between 1.5 and 2 after at most 20 iterations.

In the calculations presented here, we have taken $M_{\star} =
1$~M$_{\sun}$, $T_i=10^3$~K, the optical depth of the atmosphere above
the disk surface $\tau_{ab}=10^{-2}$ and a background temperature
$T_b=10$~K.  In the optically thick regions of the disk, the value of
$H$ is independent of the value of $\tau_{ab}$ we choose.  This is not
the case in optically thin regions where we find that, as expected, the
smaller $\tau_{ab},$ the larger $H$.  However, this dependence of $H$ on
$\tau_{ab}$ has no physical significance, since the surface mass
density, the optical thickness through the disk and the midplane
temperature hardly vary with $\tau_{ab}$.  This is because the mass is
concentrated towards the disk midplane in a layer with thickness
independent of $\tau_{ab}.$

\section{Results}
\label{sec:results}

\begin{figure}
%\epsscale{.80}
\plotone{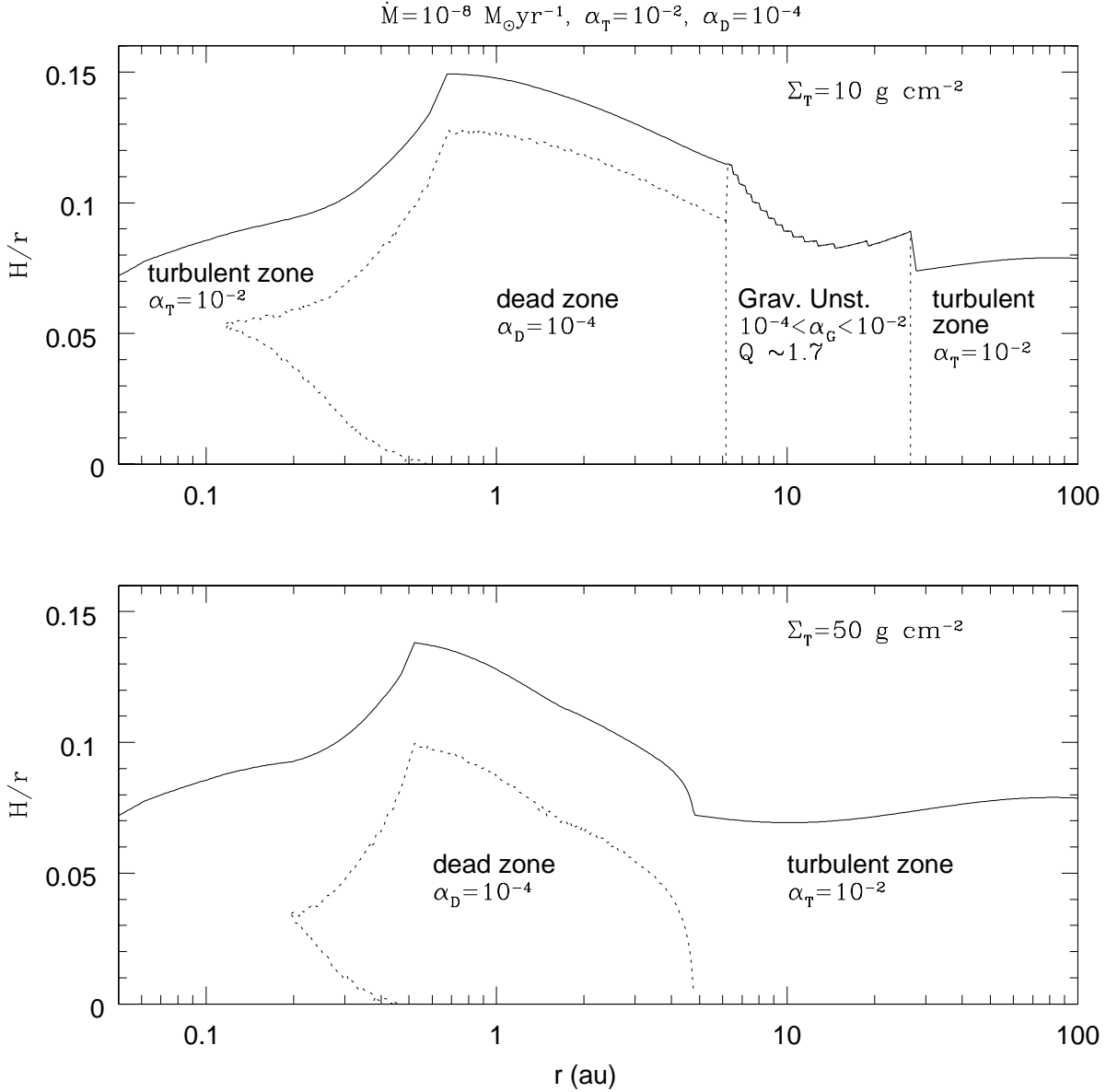} \caption{ $H/r$ versus $r$ (in au) for
$\dot{M}=10^{-8}$~M$_\sun$~yr$^{-1}$, $\alpha_T=10^{-2}$,
$\alpha_D=10^{-4}$, $T_i=10^3$~K and $\Sigma_T=10$~g~cm$^{-2}$ ({\em
upper plot}) and 50~g~cm$^{-2}$ ({\em lower plot}).  The solid line
delimits the surface of the disk whereas the dotted lines delimit the
different regions in the disk. For $\Sigma_T=10$~g~cm$^{-2}$, there is a
dead zone that extends from $r \sim 0.1$~au to $r \sim 6$~au and the
disk is gravitationally unstable from $r \sim 6$~au up to $r \sim
27$~au.  For $\Sigma_T=50$~g~cm$^{-2}$, the dead zone extends from $r
\sim 0.2$~au to $r \sim 5$~au and there is no gravitationally unstable
region.  \label{fig1}}
\end{figure}

In figure~\ref{fig1}, we plot $H/r$ versus $r$ (from 0.05 to 100~AU) for
$\dot{M}=10^{-8}$~M$_\sun$~yr$^{-1}$, $\alpha_T=10^{-2}$,
$\alpha_D=10^{-4}$ and $\Sigma_T=10$ and 50~g~cm$^{-2}$.  
The solid line
delimits the surface of the disk whereas the dotted lines delimit the
different regions in the disk.  For $\Sigma_T=10$~g~cm$^{-2}$, there is
a dead zone that extends from about 0.1~AU to about 6~AU.  Note that at
small radii, this dead zone is present only at some intermediate values
of $z$, although the ionizing photons cannot penetrate down to the disk
midplane.  This is because, if the disk starts turbulent there, the
temperature near the midplane is large enough for the gas to be
thermally ionized and therefore to remain turbulent.  And if the
disk does not start turbulent there, the temperature in the dead zone is
large enough for thermal ionization to be above the citical level for
the onset of the MRI, so that any perturbation will lead to turbulence.
From $r \sim 6$~AU up to about 27~AU, the disk is gravitationally
unstable. In this region, $\alpha_G$ increases with $r$ from $10^{-4}$
up to $10^{-2}$.  The disk is indeed cooler in the outer parts so that,
if $\alpha_G$ were kept constant, $Q$ would decrease when $r$ increases.
Therefore, to maintain $Q$ constant, $\alpha_G$ has to increase with
radius.  The inner parts of the disk, where the temperature is larger
than $T_i$, and the surface and outer parts, where the ionizing photons
can penetrate, are turbulent.  For $\Sigma_T=50$~g~cm$^{-2}$, the dead
zone extends from $r \sim 0.2$~AU to $r \sim 5$~AU and there is no
gravitationally unstable region.  This is because the dead zone is less
extended and less massive than for smaller values of $\Sigma_T$.  As the
ionizing photons can penetrate deeper than in the previous case, the
turbulent zone is much more extended.

In this model, there is a range of radii where the mass of the dead zone
is much larger than that of the active layer.  We denote $\Sigma_a$ and
$\Sigma_d$ the column density of the active layer and dead zone,
respectively.  At $r=1$~AU, where the dead zone is vertically very
extended, we have $\Sigma_a/\Sigma_d = 0.6 \times 10^{-2}$ and $4 \times
10^{-2}$ for $\Sigma_T=10$ and 50~g~cm$^{-2}$, respectively.  Whether
hydrodynamic fluctuations driven by the turbulence in the active layers
can penetrate down to the disk midplane in that case is not clear.

\begin{figure}
%\epsscale{.80}
\plotone{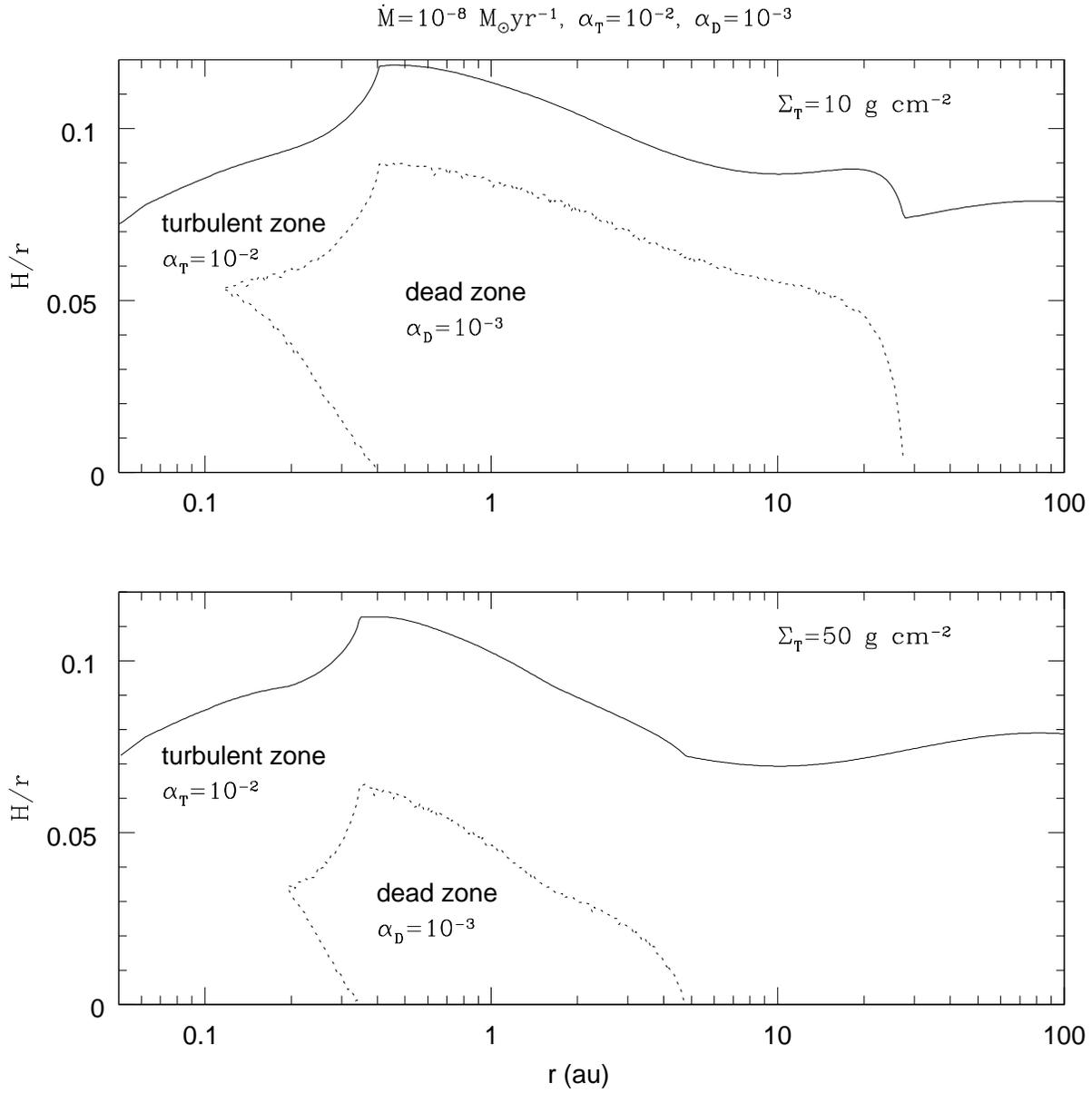} \caption{ Same as Fig.~\ref{fig1} but for
$\alpha_T=10^{-2}$ and $\alpha_D=10^{-3}$. For
$\Sigma_T=10$~g~cm$^{-2}$, the dead zone extends from $r \sim 0.1$~au to
$r \sim 27$~au whereas for $\Sigma_T=50$~g~cm$^{-2}$, it extends from $r
\sim 0.2$~au to $r \sim 5$~au.  There is no gravitationally unstable
region in that case and the vertical extent of the dead zone is smaller
than for $\alpha_D=10^{-4}$.  \label{fig2}}
\end{figure}

In order to get a model with higher values of $\Sigma_a/\Sigma_d$, we
have investigated the case $\alpha_T=10^{-2}$ and $\alpha_D=10^{-3}$
($\dot{M}=10^{-8}$~M$_\sun$~yr$^{-1}$ being the same as above).  This is
close to the values found by Fleming \& Stone~(2003) in their
simulations in which $\Sigma_a/\Sigma_d$ was varied between 0.187 and
0.75.  Figure~\ref{fig2} is the same as figure~\ref{fig1} but with this
larger value of $\alpha_D$.  
We note that the range of radii at which
the disk is not fully turbulent is the same as before.  Here however,
there is no gravitationally unstable region.  The dead zone being less
massive, it remains gravitationally stable.  For
$\Sigma_T=10$~g~cm$^{-2}$, the dead zone now extends from $r \sim
0.1$~AU up to about 27~AU.  The vertical extent of the dead zone is
smaller than in the case $\alpha_D=10^{-4}$.  At $r=0.4$~AU, where the
dead zone is the more extended, $\Sigma_a/\Sigma_d = 10^{-2}$ and $0.1$
for $\Sigma_T=10$ and 50~g~cm$^{-2}$, respectively.  This latter value
is closer to those investigated by Fleming \& Stone~(2003).  Note that
if we were to assume that the Reynolds stress in the midplane were 10\%
of the Maxwell stress at the boundary of the dead/active zone, as found
by Fleming~\& Stone~(2003), we would get $\alpha_D=0.03 \alpha_T$ after
normalizing the stress by the thermal pressure at these locations.  But
of course the angular momentum flux is spread throughout the different
zones and in principle $\alpha$ should vary with $z$.

\begin{figure}
%\epsscale{.80}
\plotone{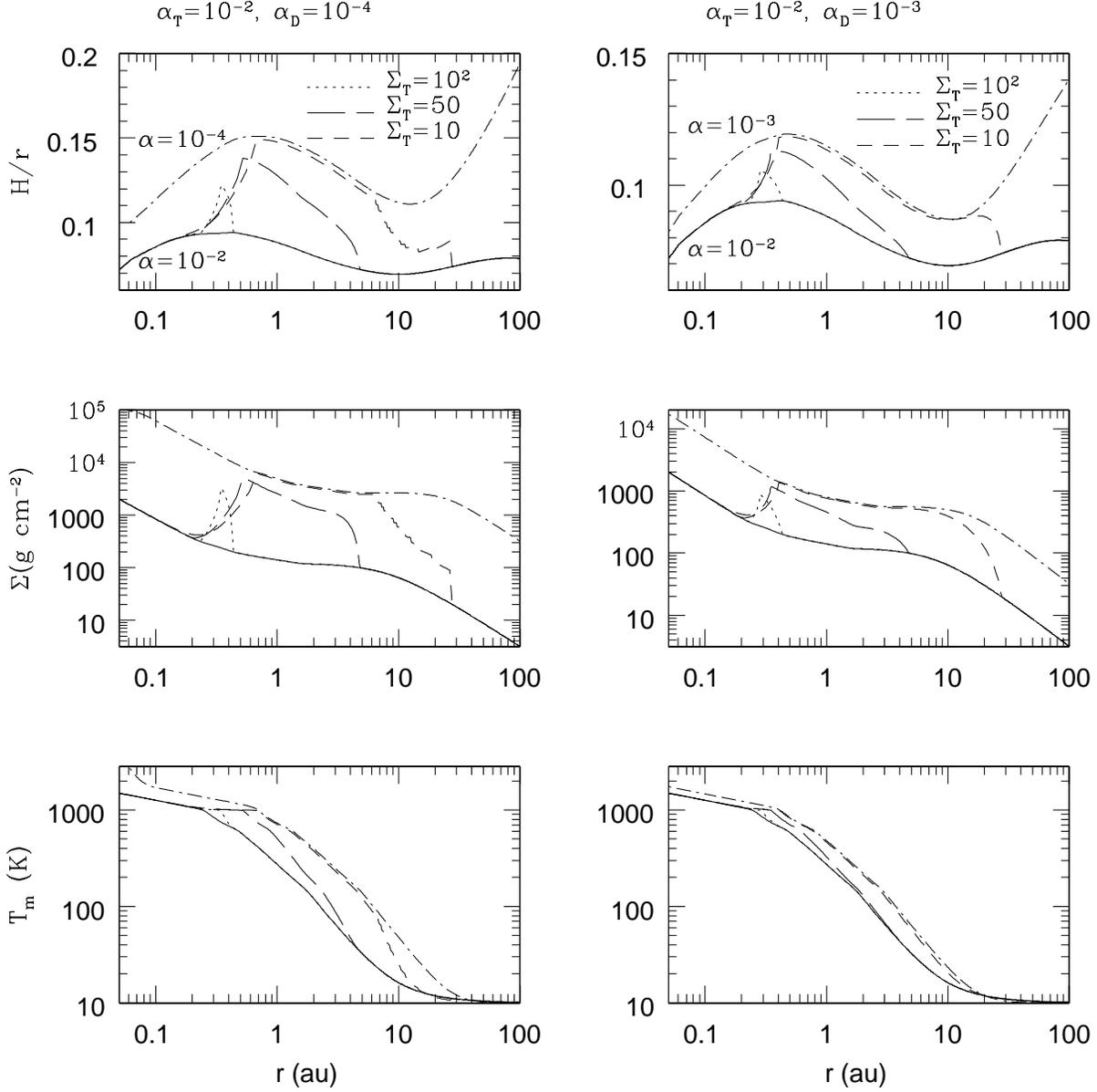} \caption{$H/r$ ({\em upper plots}), $\Sigma$ in
g~cm$^{-2}$({\em middle plots}) and midplane temperature $T_m$ in K
({\em lower plots}) versus $r$ (in au) for
$\dot{M}=10^{-8}$~M$_\sun$~yr$^{-1}$, $\alpha_T=10^{-2}$ and
$\alpha_D=10^{-4}$ ({\em left plots}) and $\alpha_T=10^{-2}$ and
$\alpha_D=10^{-3}$ ({\em right plots}).  The solid and dotted--dashed
curves correspond to standard disk models (no dead zones) with constant
$\alpha=\alpha_T$ and $\alpha=\alpha_D$, respectively.  The dotted,
long--dashed and short--dashed curves correspond to the disk models with
$\Sigma_T=10^2$, 50 and 10~g~cm$^{-2}$, respectively.  In the parts
where there is a dead zone, the disk is thicker, more massive and
hotter. \label{fig3}}
\end{figure}

In figure~\ref{fig3}, we plot $H/r$, $\Sigma$ and the midplane
temperature $T_m$ versus $r$ for the same parameters as in
figures~\ref{fig1} and~\ref{fig2}.  
For comparison, we also plot the
curves corresponding to standard disk models (with no dead zone and no
adjustment in the gravitationally unstable parts) with constant
$\alpha=\alpha_T$ and $\alpha=\alpha_D$.  As expected, in the parts
where there is a dead zone, the disk is thicker, more massive and
hotter.  This is because the vertically averaged value of $\alpha$
decreases as the vertical extent of the dead zone increases, and $H/r$,
$\Sigma$ and $T_m$ increase as $\alpha$ decreases (e.g., Frank et
al. 1992).  We see that for low values of $\Sigma_T$, i.e. when the
ionizing photons do not penetrate deep into the disk, the turbulent
layer is thin and the properties of the disk at the radii where there is
a dead zone are very close to that of a standard disk with constant
$\alpha=\alpha_D$.  For all the values of $\Sigma_T$, $H/r$ has a
maximum at a fraction of or around 1~AU, which produces a puffing up of
the disk.  This is because the mass density is very high there, so the
dead zone is vertically extended, which makes the disk hotter.  Note
that a maximum is also present in the standard disk models with constant
$\alpha$.  The profiles of $H/r$, $\Sigma$ and $T_m$ tend to be a bit
steeper than in standard disk models with no dead zone, as they are
bracketed by the profiles corresponding to constant $\alpha=\alpha_T$
and $\alpha=\alpha_D$ models. 

\begin{figure}
%\epsscale{.80}
\plotone{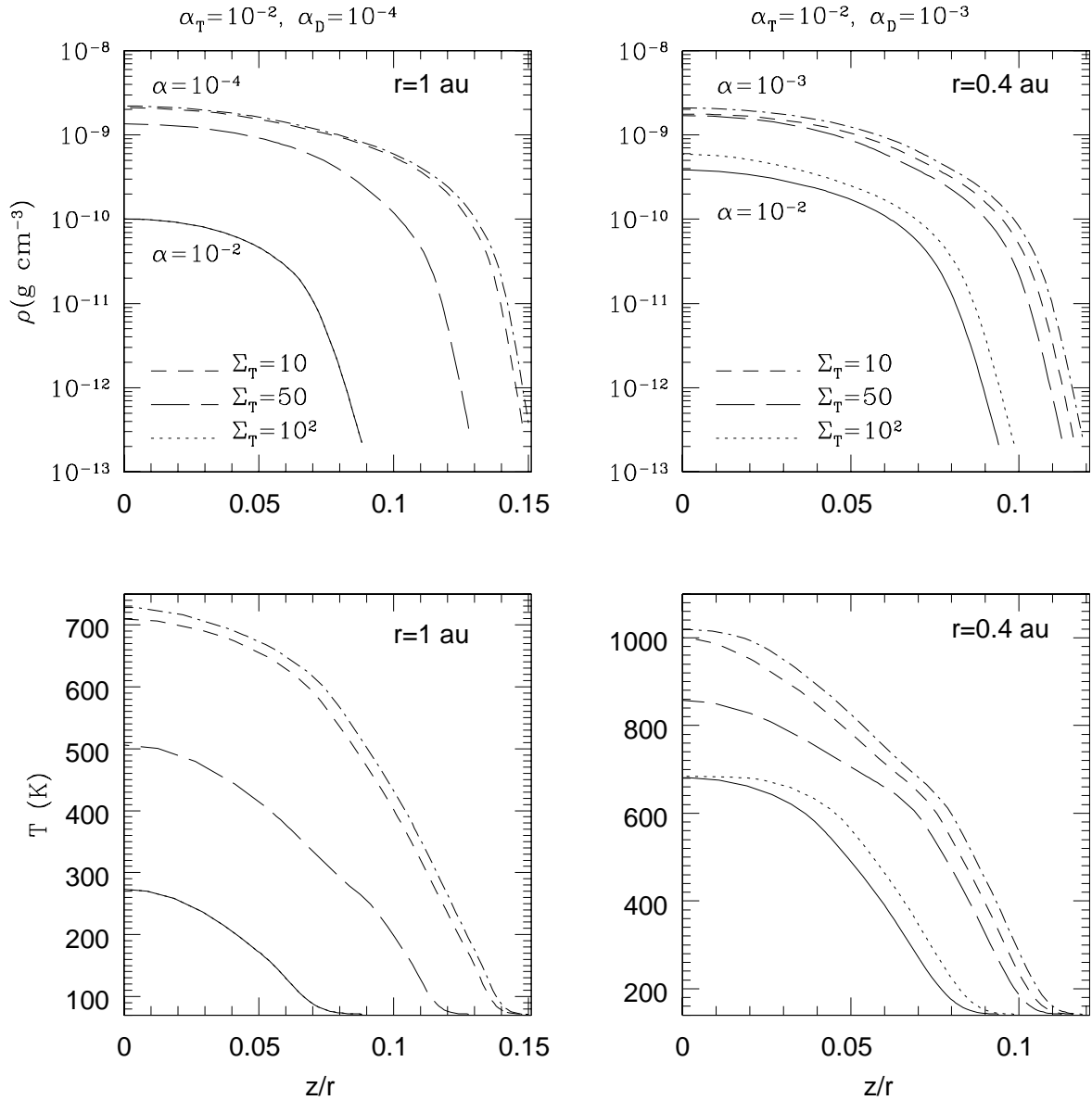} \caption{ $\rho$ in g cm$^{-3}$ ({\em upper plots})
and $T$ in K ({\em lower plots}) as a function of $z/r$ for
$\dot{M}=10^{-8}$~M$_\sun$~yr$^{-1}$, $\alpha_T=10^{-2}$ and
$\alpha_D=10^{-4}$ at $r=1$~au ({\em left plots}) and $\alpha_T=10^{-2}$
and $\alpha_D=10^{-3}$ at $r=0.4$~au ({\em right plots}).  The solid and
dotted--dashed curves correspond to standard disk models (no dead zones)
with constant $\alpha=\alpha_T$ and $\alpha=\alpha_D$, respectively.
The dotted, long--dashed and short--dashed curves correspond to the disk
models with $\Sigma_T=10^2$, 50 and 10~g~cm$^{-2}$, respectively.  On
the left plots, the solid and dotted lines are indistinguishable.  The
vertical structure of the disk depends on the vertical extent of the
dead zone.  \label{fig4}}
\end{figure}

In figure~\ref{fig4}, we plot $\rho$ and $T$ as a function of $z/r$ at
$r=1$~AU for $\alpha_D=10^{-4}$ and at $r=0.4$~AU for $\alpha_D=10^{-3}$
for the same parameters as in figure~\ref{fig3}.  When the dead zone
extends down to the disk midplane, the disk structure is very close to
that of a standard disk with constant $\alpha=\alpha_D$ for the lowest
value of $\Sigma_T$.  For the largest values of $\Sigma_T$, the dead
zone is small and the disk is similar to a standard disk with constant
$\alpha=\alpha_T$.  For $\alpha_D=10^{-4}$, we actually cannot
distinguish between the two models for $\Sigma_T=10^2$~g~cm$^{-2}$.  We
see that the mass density in the dead zone is typically 10 times larger
than that of a standard fully turbulent disk with constant
$\alpha=\alpha_T$.

\begin{figure}
%\epsscale{.80}
\plotone{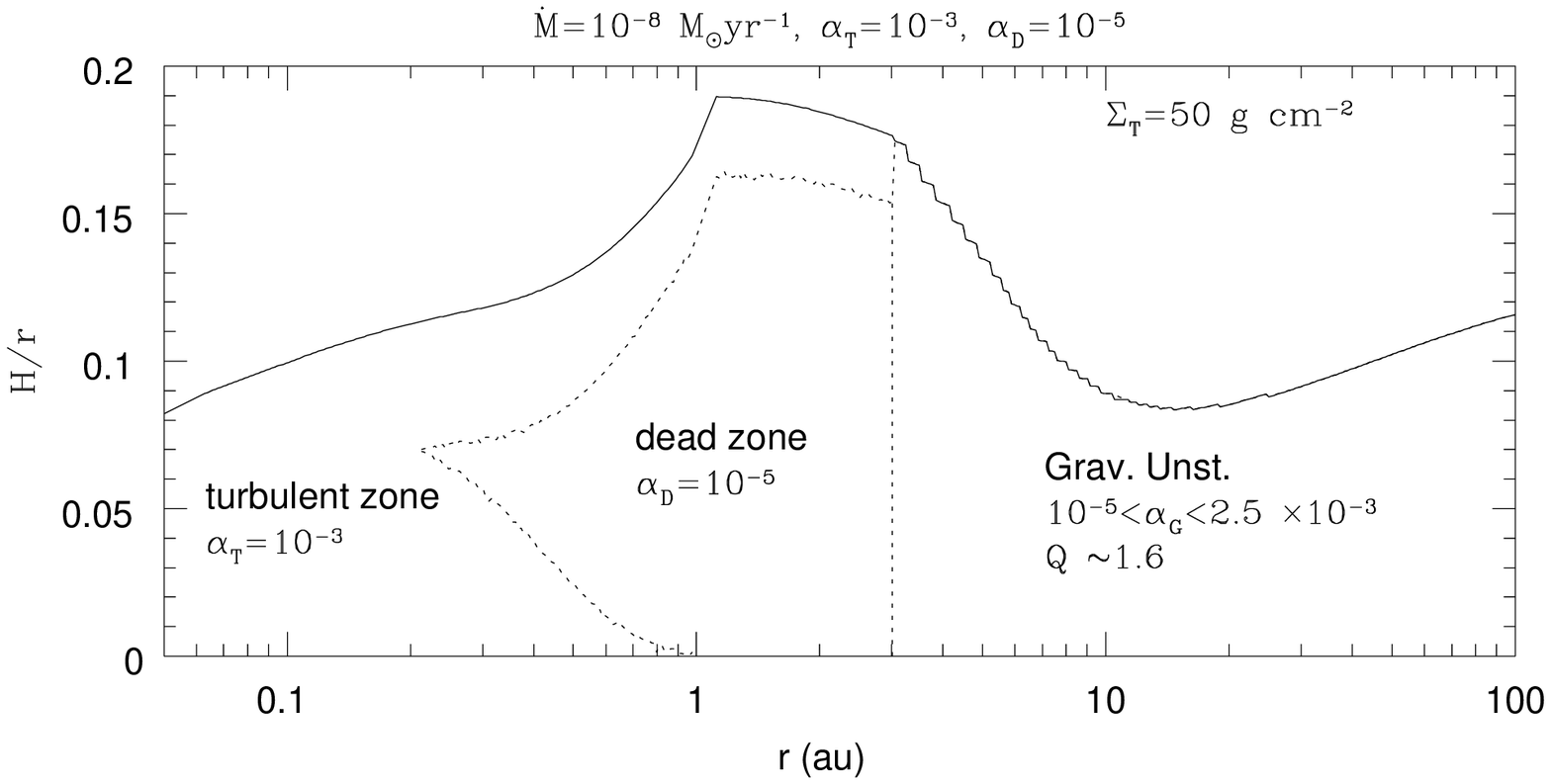} \caption{Same as Fig.~\ref{fig1} but for
$\alpha_T=10^{-3}$, $\alpha_D=10^{-5}$ and $\Sigma=50$~g~cm$^{-2}$.
 Here the disk is gravitationally unstable from $r\sim 3$~au all the way
 up to the outer radius.
\label{fig5}}
\end{figure}

In figure~\ref{fig5}, we plot $H/r$ versus $r$ for
$\dot{M}=10^{-8}$~M$_\sun$~yr$^{-1}$, $\alpha_T=10^{-3}$,
$\alpha_D=10^{-5}$ and $\Sigma_T=50$~g~cm$^{-2}$, i.e. the same values
as above except for lower values of $\alpha_T$ and $\alpha_D$.  We
checked that decreasing $\alpha_T$ and $\alpha_D$ by a factor of ten has
almost exactly the same effect as increasing $\dot{M}$ by the same
factor, all the other parameters being kept fixed.  When $\alpha_T$ and
$\alpha_D$ are smaller or $\dot{M}$ is larger, the disk is more massive
and therefore gravitational instabilities develop in a more extended
region, up to the disk outer boundary.  To keep $Q$ constant $\sim 1.6$
there, $\alpha_G$ has to reach $2.5 \times 10^{-3}$ in the disk outer
parts, i.e. to become larger than $\alpha_T$.

We have also run models with $\alpha_T=10^{-2}$ and $\alpha_D=10^{-3}
\alpha_T$.  We get similar results in that case, with the locations
where there is a dead zone being hotter, thicker and more massive.

We note that in all the models presented above, the puffing up of
the inner parts of the disk casts a shadow over the outer parts, which
therefore do not reprocess stellar radiation.  The models presented here
are self--consistent in a sense that if the disk is not illuminated by
the star to begin with, it will not be able to reprocess the stellar
radiation at any time.  There may however exist another type of
solutions in which stellar irradiation is important.  Indeed, if the
outer parts of the disk are illuminated at some point (before a
steady--state is reached for instance), a flaring is produced that could
be maintained.  To test this hypothesis, we have modified the surface
temperature.  Instead of using equation~(\ref{Ts}) to calculate $T_s$,
we fix $T_s = 300 \; \left( r / {\rm AU} \right)^{-1/2}$~K.  This
corresponds to $T_s \simeq 950$~K, 300~K and $\simeq 95$~K at $r=0.1$, 1
and 10~AU, respectively. These values are close to the surface
temperatures computed by D'Alessio et al. (1999) in irradiated disk
models for $\dot{M}=10^{-8}$~M$_\sun$~yr$^{-1}$ and $\alpha=10^{-3}$,
this value of $\alpha$ being intermediate between $\alpha_T$ and
$\alpha_D$.  With this new $T_s$, the profile of $H/r$ we get still
shows a shadowing of the outer parts.  This suggests that even if the
disk outer parts are illuminated by the central star at some point,
before a steady--state is reached, they cannot flare enough to remain
irradiated after a steady--state is established, so that reprocessing of
stellar radiation does not play a role in the steady--state models.

\section{Discussion and conclusion}
\label{sec:discussion}

In this paper, we have considered a disk in which transport is produced
by either: i) Maxwell stress resulting from MHD turbulence in the
(gravitationally stable) regions where the gas couples well to the
magnetic field; ii) gravitational stress in the parts which are
gravitationally unstable; iii) Reynolds stress associated with
hydromagnetic waves driven by the turbulence in adjacent layers and
Maxwell stress associated with large scale field siphoned off from these
turbulent layers in the parts which are not turbulent and are
gravitationally stable (dead zones).  We have modeled the transport
using the Shakura \& Sunyaev prescription, with $\alpha_D$ being 10 to
$10^3$ times smaller than $\alpha_T$, and $\alpha_G$ being adjusted such
as to give a Toomre parameter $Q \sim 1.5$ in the gravitationally
unstable regions.  Although such a modeling is in principle not valid
when the transport is non--local, we believe it does not invalidate the
gross features of the models we have presented.

We have found that steady--state models of such a disk exist, and
that they are physically reasonable.  
%For a given $\dot{M}$ and at a given radius, if we define a
%vertically--averaged value of $\alpha$, then the structure is basically
%that of a standard disk with no dead zone and characterized by this
%constant value of $\alpha$.  
Since $\dot{M} = 3 \pi \langle \nu \rangle \Sigma \propto \alpha c_s H
\Sigma$, mass flow through the disk can be uniform only if the disk is
more massive, hotter and thicker at the radii where the
vertically--averaged value of $\alpha$ is smaller, i.e. where there is a
dead zone.  Note that even though the disk is thicker at the locations
where there is a dead zone, it remains thin for the parameters
investigated here.  In disks in which the dead zone tends to be massive,
which is the case for the lowest values of $\alpha_D$ and/or the lowest
values of $\Sigma_T$ investigated here, gravitational instabilities
control the dynamics of part of the disk.

Whether these models are realistic or not depends on whether
hydrodynamical fluctuations driven by the turbulent layers can penetrate
all the way inside the dead zone.  This may be more easily achieved when
the ratio of the mass of the dead zone to that of the active layer is
the smallest, which in our models corresponds to
$\alpha_D/\alpha_T=0.1$.

If the disk is at some stage of its evolution out of steady--state, then
the surface density will change in such a way as for the disk to evolve
toward steady--state.  To see this, let us consider a disk in which
$\dot{M}$ decreases inward at some location. Then mass will accumulate
there as accretion through this region is slower, until $\Sigma$ is
large enough for the flow to be steady.  In contrast, if $\dot{M}$
increases inward at some location, accretion there will be faster and
mass will be depleted until $\Sigma$ is reduced to the level where a
steady--state is reached.  The timescale $t_{\nu}$ for establishing
steady--state at the radii where there is a dead zone may be long
though, as it is given by:

\begin{displaymath}
t_{\nu} = \frac{r^2}{3 \langle \nu \rangle} \sim \frac{1}{3 \langle 
\alpha \rangle} \left( \frac{r}{H}  \right)^2 \Omega^{-1} ,
\end{displaymath}

\noindent where $\langle \alpha \rangle = \int_{-H}^H \rho \alpha dz/
\Sigma$.  With $\langle \alpha \rangle = 10^{-2}$ and $H/r=0.1$, we get
$t_{\nu} \sim 5 \times 10^2$ and $6 \times 10^3$~years at 1 and 5~AU,
respectively, which is much smaller than the disk lifetime.  However, if
the dead zone at these radii is vertically very extended, we have
$\langle \alpha \rangle \sim \alpha_D$, so that the viscous timescale
can get 100 times longer if $\alpha_D=10^{-4}$.  At 1~AU we still get a
timescale significantly smaller than the disk lifetime, but this is not
the case at 5~AU, where the disk may not be able to reach steady--state.
If these parts of the disk build--up from mass inflowing from further
out, then the mass in the dead zone will slowly increase.  This process
will never produce outbursts as envisioned by Gammie~(1999) though.
Indeed, before enough mass could pile--up for an outburst to be
produced, a steady--state would be achieved in which mass would be
transported either by density waves or Reynolds stress driven by the
turbulent layers.

Note that the spectral energy distribution of a steady disk such as
those considered in this paper is not different from that of a standard
disk with no dead zone, as the total flux emitted at the surface depends
only on $\dot{M}$.  The surface temperature also does not depend on
$\Sigma_T$, as can be seen from equation~(\ref{Ts}).  However, the
temperature just below the surface is significantly larger in disks with
dead zones than in standard disks.  The chemistry there would therefore
be different.  Grain properties may also differ.  To study this,
irradiation of the disk by the central star would have to be taken into
account.  Although it will not be important beyond the regions where
there is a dead zone, as these parts are shielded from the stellar
radiation due to the puffing up of the disk, it will be important in the
parts where $H/r$ increases.

Finally, we comment that a steady--state disk with a dead zone is a more
favorable environment for planet formation than a standard disk, as the
dead zone, which in general encompasses the region of planet formation,
is significantly more massive.  Planet formation would therefore be
faster there (Lissauer 1993).  Note however that there is an issue as
how to stop type~I migration of the cores in these dead zones.
Turbulence, which has been shown to alter this type of migration (Nelson
\& Papaloizou 2004), cannot be invoked.  But if large scale fields are
siphoned off in these regions from the turbulent layers, they may
prevent the cores from migrating too fast (Terquem 2003, Fromang et
al. 2005, Muto et al. 2008).  Planetary cores may also be stopped at the
interface between the dead zone and the inner turbulent region, where
the surface mass density varies sharply, as suggested by Masset et
al. (2006).  This process however relies on the (positive) corotation
torque becoming dominant over the Lindblad torque due to the density
gradient and was studied in the context of a laminar viscous disk.  It
is not clear how the corotation torque is affected by the turbulence at
the transition between the active and dead zones, and whether it would
still be as efficient in this context.

\acknowledgments

I am grateful to John Papaloizou and Steven Balbus for their comments on
an early draft of this paper which led to significant improvements.
Part of this work was performed while the author was attending the
program {\em Star Formation Through Cosmic Time} at the Kavli Institute
for Theoretical Physics at UCSB, funded by the US National Science
Foundation under grant PHY05--51164.  I thank the staff and scholars at
KITP for their hospitality.

\clearpage

%% the file, uncomment the \plotone, \plottwo, and
%% If you need a layout that cannot be achieved with \plotone or
%% \plottwo, you can invoke the graphicx package directly with the
%% \includegraphics command or use \plotfiddle. For more information,
%% please see the tutorial on "Using Electronic Art with AASTeX" in the
%% documentation section at the AASTeX Web site,
%% http://www.journals.uchicago.edu/AAS/AASTeX.
%%
%% The examples below also include sample markup for submission of
%% supplemental electronic materials. As always, be sure to check
%% the instructions to authors for the journal you are submitting to
%% for specific submissions guidelines as they vary from

%% This example uses \plotone to include an EPS file scaled to
%% 80% of its natural size with \epsscale. Its caption
%% has been written to indicate that additional figure parts will be
%% available in the electronic journal.

\onecolumn

\clearpage

\clearpage

\clearpage

\clearpage

%\clearpage

\begin{thebibliography}{}

\bibitem[armitage01]{armitage01} Armitage, P. J., Livio M., \& Pringle,
		J. E. 2001, \mnras,

\bibitem[balbus91]{balbus91} Balbus, S. A., Hawley, J. F. 1991, \apj,
		376, 214

\bibitem[balbus00]{balbus00} Balbus, S. A., \& Hawley, J. F. 2000, 
		From Dust to Terrestrial Planets,  W. Benz,
		R. Kallenbach \& G. W. Lugmair, ISSI Space Sciences
		Series 9, Kluwer, p.~39

\bibitem[balbus99]{balbus99} Balbus, S. A., \& Papaloizou,
		J. C. B. 1999, \apj, 521, 650

\bibitem[bell94]{bell94} Bell, K. R., \& Lin, D.N.C. 1994, ApJ, 427, 987

\bibitem[brandenburg95]{brandenburg95} Brandenburg, A., Nordlund, A.,
		Stein, R., Torkelsson, U. 1995, \apj, 446, 741

\bibitem[dalessio98]{dalessio98} D'Alessio, P., Canto, J., Calvet, N.,
		Lizano, S. 1998, \apj, 500, 411

\bibitem[dalessio99]{dalessio99} D'Alessio, P., NURIA CALVET,2 LEE HARTMANN,2 SUSANA LIZANO,1 AND JORGE CANTO1
                                                             


\bibitem[fleming03]{fleming03} Fleming, T. P., Stone, J. M., \& Hawley,
		J. F., 2000 \apj, 530, 464

\bibitem[frank92]{frank92} Frank, J., Kimg, A., \& Raine, D. 1992,
		Accretion Power in Astrophysics, Cambridge Astrophysics
		Series

\bibitem[fromang02]{fromang02} Fromang, S., Terquem, C., \& Balbus,
		S. A. 2002, \mnras, 329, 18

\bibitem[fromang05]{fromang05} Fromang, S., Terquem, C., \& Nelson,
		R. P. 2005, \mnras, 363, 943

\bibitem[fromang06]{fromang06} Fromang, S., \& Nelson, R. P. 2006, A\&A,
		457, 343

\bibitem[fromang07]{fromang07} Fromang, S., Papaloizou, J., Lesur, G.,
		\& Heinemann, T. 2007, A\&A, 476, 1123

\bibitem[gammie96]{gammie96} Gammie, C. F. 1996, \apj, 457, 355

\bibitem[gammie99]{gammie99} Gammie, C. F. 1999, Astrophysical Discs,
		J. A. Sellwood \& J. Goodman, ASP Conf. Series,
		160, p.~122

\bibitem[hawley95]{hawley95} Hawley, J. F.,Gammie, C. F., Balbus, S. A.
		1995, \apj, 440, 742

\bibitem[igea99]{igea99} Igea, J., \& Glassgold, A. E. 1999, \apj, 518, 848

\bibitem[kenyon]{kenyon87} Kenyon, S. J., Hartmann, L. 1987, \apj, 323, 714 

\bibitem[laughlin94]{laughlin94} Laughlin, G., \& Bodenheimer, P. 1994,
		\apj, 436, 335

\bibitem[lesur07]{lesur07} Lesur, G., \& Longaretti, P.--Y. 2007,
		\mnras, 378, 1471

\bibitem[lissauer93]{lissauer93} Lissauer, J. J. 1993, ARA\&A, 31, 129

\bibitem[lodato04]{lodato04} Lodato, G., \& Rice, W. K. M. 2004, /mnras,
		351, 630

\bibitem[matsumura03]{matsumura03} Matsumura, S., Pudritz, R. E. 2003,
		\apj, 598, 645

\bibitem[matsumura06]{matsumura06} Matsumura, S., Pudritz, R. E. 2003,
		\mnras, 365, 572

\bibitem[masset06]{masset06} Masset, F. S., Morbidelli, A., Crida, A.,
		\& Ferreira, J. 2006, \apj, 642, 478

\bibitem[nelson04]{nelson04} Nelson, R. P., \& Papaloizou,
		J. C. B. 2004, \mnras, 350, 849

\bibitem[muto08]{muto08} Muto, T., Machida, M. N., \& Inutsuka, S.--I.,
		2008, arXiv:0712.1060v2 [astro--ph]

\bibitem[papaloizou99]{papaloizou99} Papaloizou, J. C. B., \& Terquem,
		C. 1999, \apj, 521, 823

\bibitem[sano00]{sano00} Sano, T., Miyama, S. M., Umebayashi, T., \&
		Nakano, T. 2000, \apj, 543, 486

\bibitem[shakura73]{shakura73} Shakura, N. I., \& Sunyaev, R. A. 1973,
		A\&A, 24, 337

\bibitem[terquem03]{terquem03} Terquem, C. E. J. M. L. J. 2003, \mnras,
		341, 1157

\bibitem[turner08]{turner08} Turner, N. J., \& Sano, T. 2008,
		arXiv:0804.2916v1 [astro--ph]



\end{thebibliography}
\end{document}